\documentclass[12pt, letterpaper]{extarticle}

\usepackage{helvet}
\usepackage{times}
\usepackage{comment}
\usepackage{placeins}
\usepackage{multirow}
\usepackage{rotating}
\usepackage{mathtools}
\usepackage{subfig}
\usepackage{arydshln}
\usepackage{longtable}
\usepackage{float}
\usepackage{multicol}
\usepackage{supertabular}
\usepackage{booktabs}
\usepackage[usenames,dvipsnames,svgnames,table]{xcolor}
\usepackage{longtable}
\usepackage[framemethod=TikZ]{mdframed}
\usepackage{fullpage}
\usepackage[switch]{lineno}
\usepackage{amsmath}
\usepackage{amssymb}
\usepackage{rotating}
\usepackage{array}
\usepackage{mathtools}
\usepackage[ruled]{algorithm2e}
\usepackage{algorithmic}
\usepackage{bm}
\usepackage{breqn}
\usepackage{comment}
\usepackage{enumitem}
\usepackage{graphics}
\usepackage{graphicx}
\usepackage{latexsym}
\usepackage{mathrsfs}
\usepackage{morefloats}
\usepackage{todonotes}
\usepackage{nicefrac}
\usepackage{authblk}

\title{The Impact of Informal Mentorship in Academic Collaborations}
\author[1]{Bedoor AlShebli}
\author[1]{Kinga Makovi}
\author[1*]{Talal Rahwan}

\affil[1]{\normalsize 
New York University, Abu Dhabi, UAE.}

\affil[*]{\footnotesize Corresponding author. E-mail:\ \ talal.rahwan@nyu.edu}
\date{}

\begin{document} 

\maketitle 

\begin{abstract}
Inspired by the numerous benefits of mentorship in academia, we study \textit{informal mentorship} in scientific collaborations, whereby a junior scientist is supported by multiple senior collaborators, without them necessarily having any formal supervisory roles. To this end, we analyze 2.5 million unique pairs of mentor-prot\'eg\'es spanning 9 disciplines and over a century of research, and we show that mentorship quality has a causal effect on the scientific impact of the papers written by the prot\'eg\'e post mentorship. This effect increases with the number of mentors, and persists over time, across disciplines and university ranks. The effect also increases with the academic age of the mentors until they reach 30 years of experience, after which it starts to decrease. Furthermore, we study how the gender of both the mentors and their prot\'eg\'e affect not only the impact of the prot\'eg\'e post mentorship, but also the citation gain of the mentors during the mentorship experience with their prot\'eg\'e. We find that increasing the proportion of female mentors decreases the impact of the prot\'eg\'e, while also compromising the gain of female mentors. While current policies that have been encouraging junior females to be mentored by senior females have been instrumental in retaining women in science, our findings suggest that the impact of women who remain in academia may increase by encouraging opposite-gender mentorships instead. 
\end{abstract}

\section*{Introduction}

Mentorship contributes to the advancement of individual careers \cite{Kram_1988, Allen_etal_2004, Scandura_1992} and provides continuity in organizations \cite{Singh_etal_2002, Tongand_Kram_2013}. By mentoring novices, senior members pass on the organizational culture, best practices, and inner workings of a profession. In this way, the mentor-prot\'eg\'e relationship provides the human glue that links generations within a field. Mentorship can also alleviate the barriers of entry for underrepresented minorities, such as women and people of color, thereby acting as an equalizing force \cite{Kanter_1977, Levine_Nidiffer_1996, Stephens_etal_2014, Gaule_Piacentinic_2018}. Most workplaces have shifted from the classic master-apprentice model towards a team-based model where the mentorship of junior members is distributed amongst the senior members of the team. As a result, \emph{informal mentorship}, whereby juniors receive support from senior colleagues without a formal supervisory role, has become commonplace \cite{Higgins_Kram_2001, Higgins_Thomas_2001}. Being informal makes such mentorship invisible and thus elusive to study.

Academia has seen a similar shift towards a team-based model \cite{Wuchty_etal_2007} which makes it an ideal application example to study informal mentorship. It has already proven to be an effective test bed to study a wide variety of topics, including team assembly \cite{Guimera_etal_2005}, the role of past experience \cite{Sekara_etal_2018}, individual productivity \cite{Acuna_etal_2012, Sugimoto_etal_2017, Liu_etal_2018b}, diversity \cite{AlShebli_etal_2018}, and innovation \cite{Uzzi_etal_2013}, thereby giving rise to the field of Science of Science \cite{Fortunato_etal_2018}. Importantly, academic papers provide a documented record of millions of collaborations spread over decades, along with an established measure of success, namely citation count. Motivated by this observation, our study of informal mentorship focuses on academic collaborations between junior and senior scientists, since such collaborations play an important role in shaping the junior scientist's persona, both in terms of their research focus \cite{Hirshman_etal_2016}, professional ethics and work culture \cite{Johnson_2007}. Furthermore, inspired by the expanding literature on gender equity and diversity in science \cite{Lariviere_etal_2013, Handley_etal_2015, Nielsen_etal_2017, Berenbaum_2019}, we analyze the mentorship experiences from the perspective of both female and male scientists.

Compared to previous studies on mentorship in academia \cite{Reskin_1979, Kirchmeyer_2005, Paglis_etal_2006, Malmgren_etal_2010, Chariker_etal_2017, Rossi_etal_2017, Lineard_etal_2018, Liu_etal_2018}, ours has the following advantages. First, we capture multiple informal mentor-prot\'eg\'e relationships per collaboration, rather than restricting our analysis to formal student-advisor relationships. Second, we avoid sample selectivity as well as recall and recency biases, since we analyze the actual scientific impact of collaborations rather than self-reported information. Third, we analyze thousands of journals spanning multiple scientific disciplines, rather than focusing on just a single journal or discipline. Fourth, we construct careful comparisons between millions of mentor-prot\'eg\'e pairs, allowing us to establish and quantify the causal effect of the mentorship experience on scientific careers. Finally, our study complements the literature that studies the impact of mentorship on attrition from science \cite{Blau_etal_2010}, as we consider prot\'eg\'es who remain scientifically active after the completion of their mentorship period.

\section*{Results}

We analyze 115 million scientists and 130 million papers gleaned from the Microsoft Academic Graph (MAG) dataset \cite{sinha2015overview}, which contains detailed records of scientific publications and their citation network. In addition to MAG we use other external data-generating techniques and sources to establish the gender of scientists and the rank of their affiliations (see Methods).

We distinguish between \textit{junior} and \textit{senior} scientists based on their academic age, measured by the number of years since their first publication. The junior years are those during which a scientist participates in graduate and postdoctoral training, and possibly the first few years of being a faculty member or researcher. In contrast, the senior years are those during which a scientist typically accumulates experience as a PI and transitions into a supervisory role. For any given scientist, we consider the first 7 years of their career to be their junior years, and the ones after that to be their senior years. Whenever a junior scientist publishes a paper with a senior scientist, we consider the former to be a \textit{prot\'eg\'e}, and the latter to be a \textit{mentor}, as long as they share the same discipline and US-based affiliation. Our use sample consists of 2.5 million unique mentor-prot\'eg\'e pairs, spanning nine disciplines (Biology, Chemistry, Computer Science, Economics, Engineering, Geology, Mathematics, Medicine, Psychology) and over a century of research.

We consider two alternative measures of mentorship quality.
The first is the \textit{average impact of mentors prior to mentorship}, where the prior impact of each mentor is computed as their average number of citations per annum up to the year of their first publication with the prot\'eg\'e. This reflects the success of mentors and their standing and reputation in their respective scientific communities. We refer to this measure as the \textit{big-shot experience}, as it captures how much of a ``big-shot'' the mentors of the prot\'eg\'e are. The second measure of mentorship quality that we consider is the \textit{average degree of mentors prior to mentorship}, where the degree of each mentor is calculated in the network of scientific collaborations 
up to the year of their first publication with the prot\'eg\'e.
We refer to this measure as the \textit{hub experience}, as it reflects how much of a ``hub'' each mentor is in the collaboration network. These two measures of mentorship experience take the role of independent variables in our study.

Having discussed our measures of mentorship quality, we now discuss the mentorship outcome, which we conceptualize as the impact of the prot\'eg\'e during their senior years without their mentors. We measure this outcome by calculating the average impact of all the papers that the prot\'eg\'e published post mentorship without their mentors. The impact of each such paper is calculated as the number of citations that it accumulated five years post publication, denoted by $c_5$ \cite{AlShebli_etal_2018}. Such an outcome measure allows us to assess the quality of the scholar that the prot\'eg\'e has become after the mentorship period has concluded.

We aim to establish whether mentorship quality (measured by big-shot experience or hub experience) has a causal effect on post mentorship outcomes,
and we aim to quantify this effect. To this end, we use coarsened exact matching (CEM), a technique widely used when estimating causal effects in observational studies \cite{Iacus_etal_2012}. Intuitively, CEM allows us to select a group of prot\'eg\'es who received a certain level of mentorship quality (treatment group), and match it to another group of prot\'eg\'es who received a lower level of mentorship quality (control group). Comparing the outcome of the two groups allows us to estimate the average causal effect. In more detail, for each measure of mentorship quality, we create a separate CEM where the treatment and control groups differ in terms of that measure, but resemble each other in terms of all of the following: the number of mentors, the year in which the prot\'eg\'e published their first mentored paper, the scientific discipline of the prot\'eg\'e, the gender of the prot\'eg\'e, the rank of the affiliation of the prot\'eg\'e on their first mentored publication, the number of years starting from the first year post mentorship until the prot\'eg\'e's last active year, and the average academic age of the mentors, which is computed based on the academic age of each mentor in the year of their first publication with the prot\'eg\'e. Finally, when studying the impact of the big-shot experience, we make sure that the two groups are similar in terms of the hub experience, and vice versa. 

For every independent variable, be it big-shot experience or hub experience, let $Q_i$ denote the $i^{\textnormal{th}}$ quintile of the distribution of that variable. Then, for $i\in\{1,2,3,4\}$, we build a separate CEM where the control and treatment groups are $Q_i$ and $Q_{i+1}$, respectively. The CEM results are depicted in Fig.~\ref{fig:CEM_results}. As can be seen, the big-shot experience increases the post-mentorship impact of prot\'eg\'es by up to 36\%; we refer to this as the \emph{big-shot effect}. The hub experience also increases the post-mentorship impact of prot\'eg\'es, although the increase never exceeds 7\%; we refer to this as the \emph{hub effect}. Clearly, regardless of the choice of control and treatment groups, the big-shot effect is always significantly greater than the hub effect. As such, we restrict our attention to the big-shot effect throughout the remainder of our study. 

In Fig.~\ref{fig:CEM_across}, we take a closer look at the big-shot effect, to understand how it is affected by the year in which the mentorship started, the age of the mentors, and the number of mentors. In particular, Fig.~\ref{fig:CEM_across}a analyzes the big-shot effect across the years of the prot\'eg\'e's first publication. As can be seen, the big-shot effect is increasing over the years, and nearly doubled in the past two decades. On the other hand, Fig.~\ref{fig:CEM_across}b analyzes the big-shot effect across varying ages of the mentors. This shows that, as the mentors get older, their impact on the prot\'eg\'e continues to increase until their academic age exceeds 30 years, after which the impact starts to decrease. In Fig.~\ref{fig:CEM_across}c, we analyze the big-shot effect given varying numbers of mentors. This shows that the impact increases with the number of mentors, and that the prot\'eg\'e  benefits from having more than 5 mentors during the mentorship period. Finally, in Supplementary Figures~S1, S2 and S3 we show that the big-shot effect persists regardless of the discipline, the university rank, and the gender of the prot\'eg\'e.

Next, we turn to an exploratory analysis where we investigate the post-mentorship impact of prot\'eg\'es while taking into consideration their gender as well as the gender of their mentors. To this end, let $F_i$ denote the set of prot\'eg\'es that have exactly $i$ female mentors. We take the prot\'eg\'es in $F_0$ as our baseline, and match them to those in $F_i$ for some $i>0$, while controlling for the prot\'eg\'e's big-shot experience, number of mentors, gender, discipline, affiliation rank, and the year in which they published their first mentored paper. Then, for any given $i>0$, we compute the change in the post-mentorship impact of the prot\'eg\'es in $F_i$ relative to the post-mentorship impact of those in $F_0$; we refer to this comparison by writing $F_0$ vs.~$F_i$. The outcomes of these comparisons are depicted for male prot\'eg\'es in Fig.~\ref{fig:gender_results}a, and for female prot\'eg\'es in Fig.~\ref{fig:gender_results}b. As shown in this figure, having more female mentors tends to decrease the mentorship outcome, and this decrease can reach as high as 35\%, depending on the number of mentors and the proportion of female mentors. 

So far in our analysis, we only considered the outcome of the prot\'eg\'es. However, mentors have also been shown to benefit from the mentorship experience \cite{Kram_1988}. With this in mind, we measure the gain of a mentor from a particular prot\'eg\'e as the average impact, $\langle c_5\rangle$, of the papers they wrote with that prot\'eg\'e during the mentorship period. We compare the average gain of a female mentor, $F$, against that of a male mentor, $M$, when mentoring either a female prot\'eg\'e, $f$, or a male prot\'eg\'e, $m$. More specifically, we compare mentor-prot\'eg\'e relationships of the type $(f,F)$ to those of the type $(m,F)$, where $f$ and $m$ are matched based on their discipline, affiliation rank, number of mentors, and the year in which they published their first mentored paper. Similarly, we compare relationships of the type $(f,M)$ to those of the type $(m,M)$, where $f$ and $m$ are matched as above. The results of these comparisons are presented in Fig.~\ref{fig:gender_results}c. In particular, the figure depicts the gain from mentoring a female prot\'eg\'e relative to that of mentoring a male prot\'eg\'e; these results are presented for female mentors and male mentors separately. As can be seen, by mentoring female instead of male prot\'eg\'es, the female mentors compromise their gain from mentorship, and suffer on average a loss of 18\% in citations on their mentored papers. As for male mentors, their gain does not appear to be significantly affected by taking female instead of male prot\'eg\'es.

In this paper, we studied informal mentorship in academia, whereby a junior scientist receives support from multiple senior collaborators without necessarily having a formal supervisory relationship. Having conceptualized mentorship quality in two ways---the \emph{big-shot} experience and the \emph{hub} experience---we found that both have an independent causal effect on the prot\'eg\'e's impact post mentorship without their mentors. Interestingly, the big-shot effect is larger than the hub effect, implying that the scientific impact of mentors matters more than their number of collaborators. We further analyzed the big-shot effect to understand how it is affected by the number of mentors, the age of the mentors, and the year in which the mentorship started. This analysis revealed that the effect of mentorship increases with the number of mentors, and nearly doubled over the past two decades. Furthermore, as the mentors get older, their impact on the prot\'eg\'e increases until they reach 30 year of experience, after which their impact decreases. We also showed that the big-shot effect persists regardless of the discipline, the university rank, and the gender of the prot\'eg\'e. Finally, we studied how the gender of both the mentors and their prot\'eg\'e affects not only the impact of the prot\'eg\'e, but also the gain of the mentors, measured by the citations of the papers they published with the prot\'eg\'e during the mentorship period. We found that increasing the proportion of female mentors tends to decrease the impact of prot\'eg\'es, and also found that the gain of female mentors decreases when mentoring female instead of male prot\'eg\'es.

While it has been shown that having female mentors increases the likelihood of female prot\'eg\'es staying in academia \cite{Gaule_Piacentinic_2018} and provides them with better career outcomes \cite{Blau_etal_2010}, such studies often compare prot\'eg\'es that have a female mentor to those who do not have a mentor at all, rather than to those who have a male mentor. In this paper, we addressed this limitation in the context of informal mentorship, and found that female prot\'eg\'es who remain in academia reap more benefits when mentored by males rather than females, while the benefits of mentorship accruing to senior female mentors are also higher when working with otherwise similar male prot\'eg\'es. These findings suggest that current diversity policies promoting female-female mentorships, as well-intended as they may be, could hinder the careers of women who remain in academia in unexpected ways. Female scientists, in fact, may benefit from encouraging opposite-gender mentorships instead throughout their careers. Policy makers should thus revisit first and second order consequences of diversity policies while focusing not only on retaining women in science, but also on maximizing their long term scientific impact. In addition to these perspectives, policies should also consider the benefits accrued from gender diversity within scientific teams, as well as the quality of their overall scientific output \cite{Bear_Woolley_2011}. More broadly, the goal of gender equity in science, regardless of the objective targeted by these policies, cannot, and should not be shouldered by senior female scientists alone, rather, it should be embraced by the scientific community as a whole.

\bibliographystyle{naturemag}
\bibliography{sample}

\newpage

\section*{Methods}

The data used for this study was obtained in October 2015 from the Microsoft Academic Graph (MAG) dataset 
\cite{sinha2015overview}. This data set includes records of scientific publications specifying the date of the publication, the authors' names and affiliations, the publication venue, and the keywords. It also contains a citation network in which every node represents a paper and every directed edge represents a citation. While the number of citations of any given paper is not provided explicitly, it can easily be calculated from the citation network in any given year. Additionally, every keyword in a given paper is positioned in a field-of-study hierarchy, the highest level of which is comprised of 19 scientific disciplines.

Using the information provided in the MAG dataset, we derive two key measures: the main discipline of scientists and their impact. In particular, we classify scientists into disciplines using the method proposed by \cite{AlShebli_etal_2018}. This method classifies any given scientist, $s_i$, using the keywords that are specified in the publications of $s_i$, which are themselves classified into disciplines by MAG. On the other hand, the impact of each scientist in any given year is derived from the citation network provided by MAG. In addition to the scientists' discipline and impact, we derive additional measures such as the scientists' gender, which is determined using ``genderize.io'' \cite{wais2016gender}, and the rank of each university, which is determined based on the 2018 Academic Ranking of World Universities, also known as the \textit{Shanghai ranking} \cite{shanghai:ranking:2018}.

Whenever a junior scientist (with academic age $\leq 7$) publishes a paper with a senior scientist (academic age $> 7$), we consider the former to be a \textit{prot\'eg\'e}, and the latter to be a \textit{mentor}. We consider the start of the mentorship period to be the year of the first publication of the prot\'eg\'e, and consider the end of the mentorship period to be the year when the prot\'eg\'e becomes a senior scientist. We analyze every mentor-prot\'eg\'e dyad that satisfies all of the following conditions: (i) the prot\'eg\'e has at least one publication during their senior years without a mentor; (ii) the affiliation of the prot\'eg\'e is in the United States throughout their mentorship years; (iii) the main discipline of the mentor is the same as that of the prot\'eg\'e; (iv) the mentor and the prot\'eg\'e share an affiliation on at least one publication; and (v) the prot\'eg\'es do not have a gap of 5-years or more in their publication history. As a consequence, our analysis excludes all scientists: (i) who never published any papers without their mentors post mentorship, as we cannot analyze their scientific impact in their senior years independent of their mentors; (ii) who only had solo-authored papers or collaborations with their junior peers or with seniors from other universities, as we cannot clearly establish who their mentors were; (iii) who had a gap longer than 5-years without any publications; and (iv) who only collaborated with senior scientists outside of their main discipline.

As our use sample we consider the ten disciplines in MAG that have the largest number of mentor-prot\'eg\'e pairs, namely Biology, Chemistry, Computer Science, Economics, Engineering, Geology, Mathematics, Medicine, Physics, and Psychology (see Supplementary Table~S1). From these we drop Physics since the mean number of authors per paper in this discipline is far larger than that of any other discipline (236, compared to 13 or below; see Supplementary Table~S2). As such, keeping Physics would render our approach of identifying mentor-prot\'eg\'e relationships quixotic and impractical. Furthermore, we dropped all mentor-prot\'eg\'e pairs where the gender of the prot\'eg\'e could not be established with at least 95\% certainty. Later on, when analyzing the gender of mentors, we also drop any mentors whose gender could not be established with at least 95\% certainty. 

\begin{figure}
\begin{center}
\includegraphics[scale=.45]{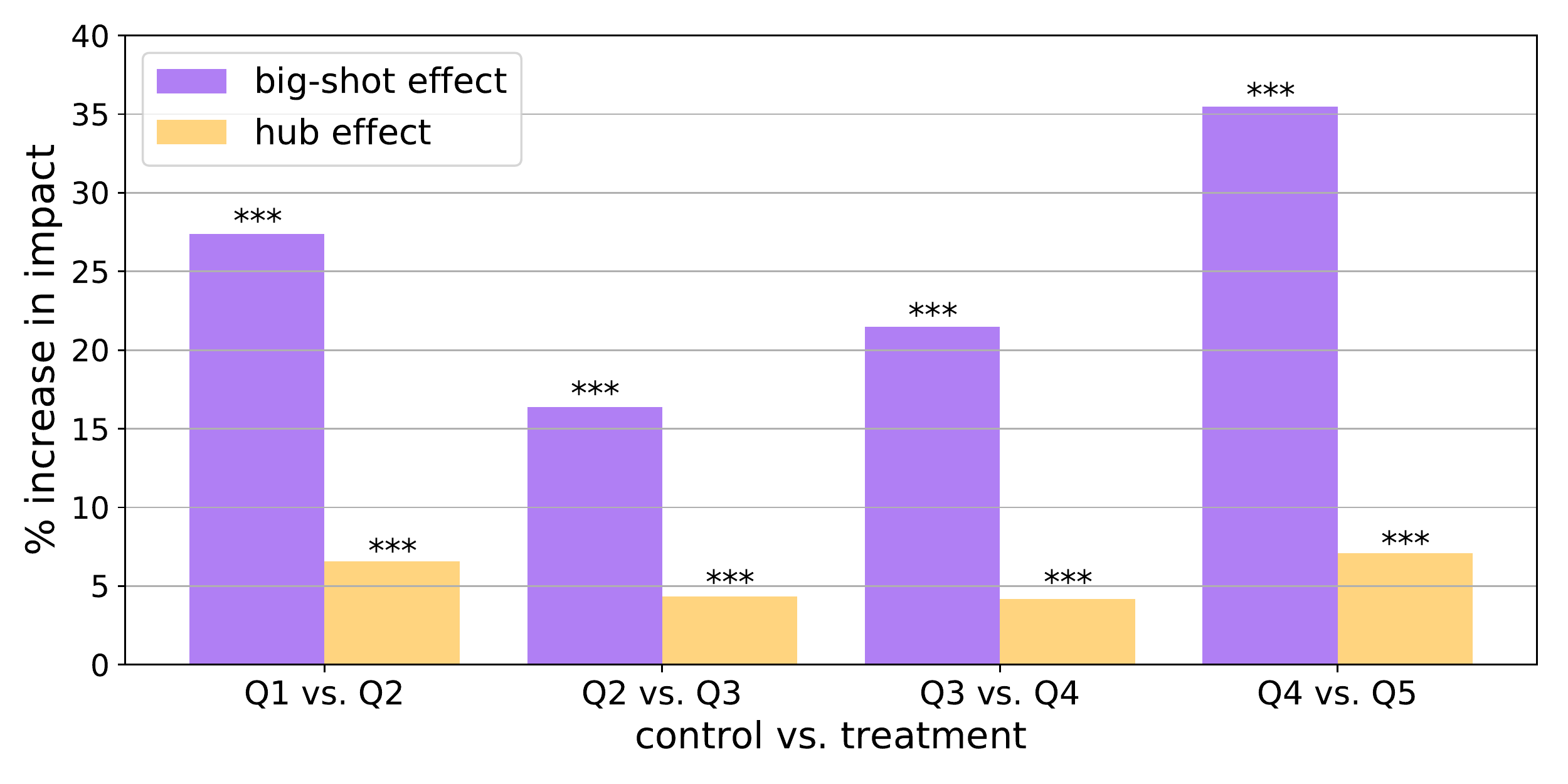} 
\caption{\textbf{The big-shot effect and hub effect.} For every independent variable, be it big-shot experience or hub experience, $Q_i$ denotes the $i^{\textnormal{th}}$ quintile of the distribution of that variable. For $i\in\{1,2,3,4\}$, we consider $Q_i$ and $Q_{i+1}$ to be the control and treatment groups, respectively, and write $Q_i$ vs.~$Q_{i+1}$ when referring to the CEM used to compare these two groups. The color of the bar indicates whether the independent variable is the big-shot experience (purple) or the hub-experience (yellow), whereas the height of the bar equals $\delta$, which is the increase in the post-mentorship impact of the treatment group relative to that of the control group; see Supplementary Tables S1 and S2 for more details. ***p$<$0.001.
}
\label{fig:CEM_results}
\end{center}
\end{figure}

\begin{figure}
\begin{center}
\includegraphics[width=\linewidth]{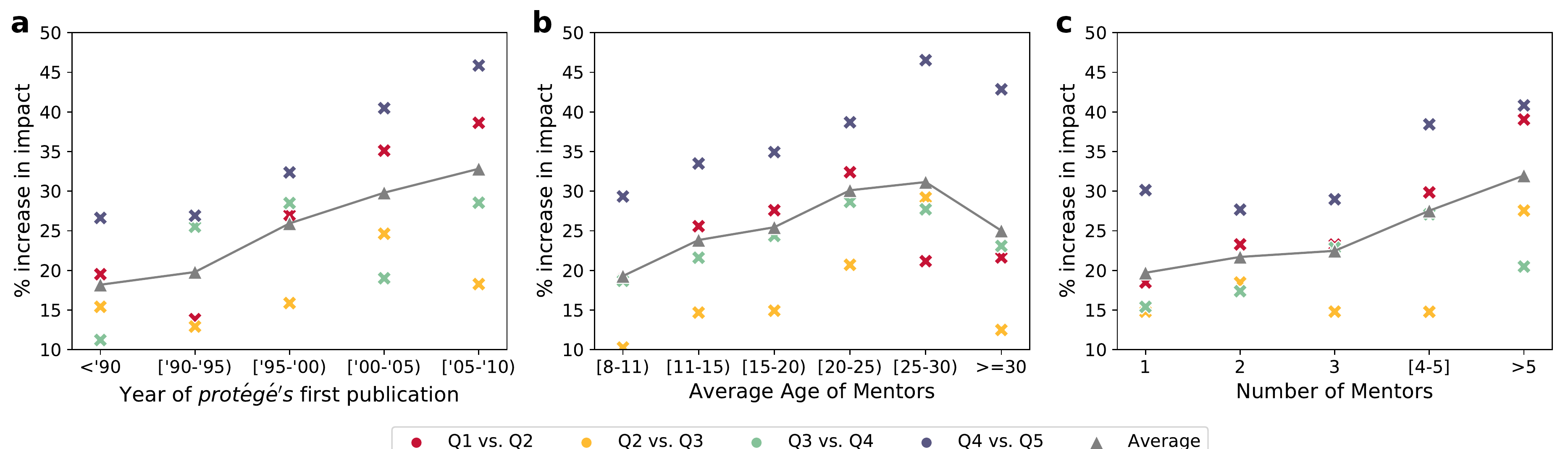} 
\caption{\textbf{Trends in big-shot effect.} \textbf{a}, Big-shot effect over time. \textbf{b}, Big-shot effect across varying ages of the mentors. \textbf{c}, Big-shot effect across different numbers of mentors. In all subfigures, $Q_i:i\in\{1,2,3,4\}$ represents the $i^{\textnormal{th}}$ quintile of the distribution of big-shot experience for all prot\'eg\'es that fall in the same bin of either the year of their first publication (subfigure~\textbf{a}), the average age of the mentors (subfigure~\textbf{b}), or the number of mentors (subfigure~\textbf{c}). Every point depicted as ``$\times$'' represents a separate CEM where the control and treatment groups are $Q_i$ and $Q_{i+1}$ for some $i\in\{1,2,3,4\}$, respectively (the color of $\times$ indicates the value of $i$). On the other hand, every point depicted as ``$\Delta$'' is the average of all the $\times$ points that correspond to the same bin on the $x$-axis. All results are statistically significant; see Supplementary Tables S5, S6 and S7 for more details on the CEMs used to produce these figures.
}
\label{fig:CEM_across}
\end{center}
\end{figure}

\begin{figure}
\begin{center}
\includegraphics[scale=0.48]{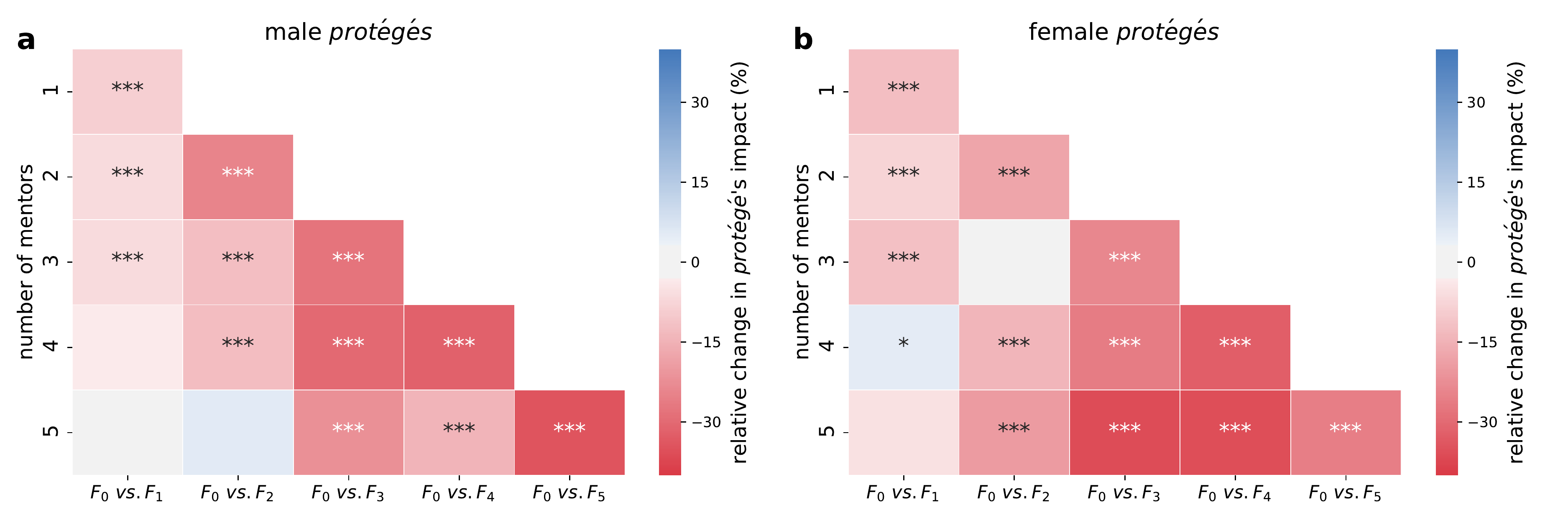} 
\includegraphics[scale=0.48]{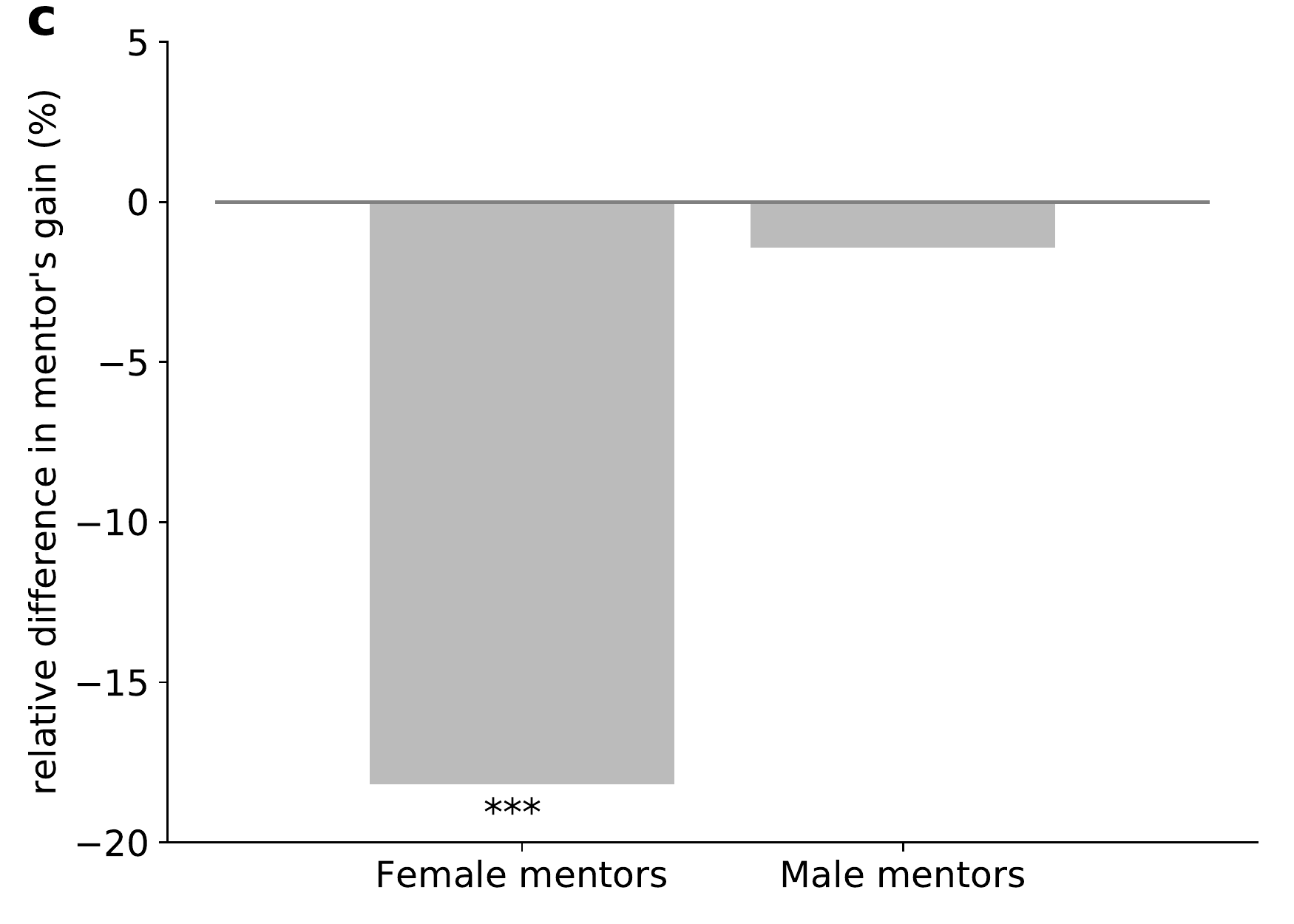} 
\caption{\textbf{The relationship between gender and the gain from mentorship.} \textbf{a}, $F_i$ denotes the set of prot\'eg\'es that have exactly $i$ female mentors. Focusing on male prot\'eg\'es, $F_0~\textnormal{vs.}~F_i:i=1,\ldots,5$ refers to the change in the post-mentorship impact of prot\'eg\'es in $F_i$ relative to the post-mentorship impact of those in $F_0$ while controlling for the prot\'eg\'e's big-shot experience, number of mentors, discipline, affiliation rank, and the year in which they published their first mentored paper. \textbf{b}, The same as (\textbf{a}) but for female prot\'eg\'es instead of male prot\'eg\'es. \textbf{c}, The gain of a mentor from a particular prot\'eg\'e is measured as the average $c_5$ of the papers they wrote with that prot\'eg\'e during the mentorship period. While controlling for the prot\'eg\'e's discipline, affiliation rank, number of mentors, and the year in which they published their first mentored paper, the figure depicts the change in the mentor's gain when mentoring a female prot\'eg\'e relative to that when mentoring a male prot\'eg\'e; results are presented for female mentors and male mentors separately. *p$<$0.05, **p$<$0.01, ***p$<$0.001.
}
\label{fig:gender_results}
\end{center}
\end{figure}

A total of 140 different Coarsened Exact Matchings (CEMs) were used to produce the results depicted in Fig.~\ref{fig:CEM_results}, Fig.~\ref{fig:CEM_across}, Fig.~S1, Fig.~S2, and Fig.~S3. Additionally, a total of 32 different matchings were used to produce the results depicted in Fig.~\ref{fig:gender_results}. More details about the confounding factors used therein, as well as the binning decisions, can all be found in the Supplementary Note~2.

\newpage

\end{document}


\renewcommand{\thetable}{S\arabic{table}}
\renewcommand{\thefigure}{S\arabic{figure}}


\baselineskip24pt

\date{\vspace{-0.1ex}}
\maketitle 
\ \\

\vspace{-1cm}
\tableofcontents

\clearpage

\section{Supplementary Figures}

\begin{figure}[H]
\centering
\includegraphics[width=\linewidth]{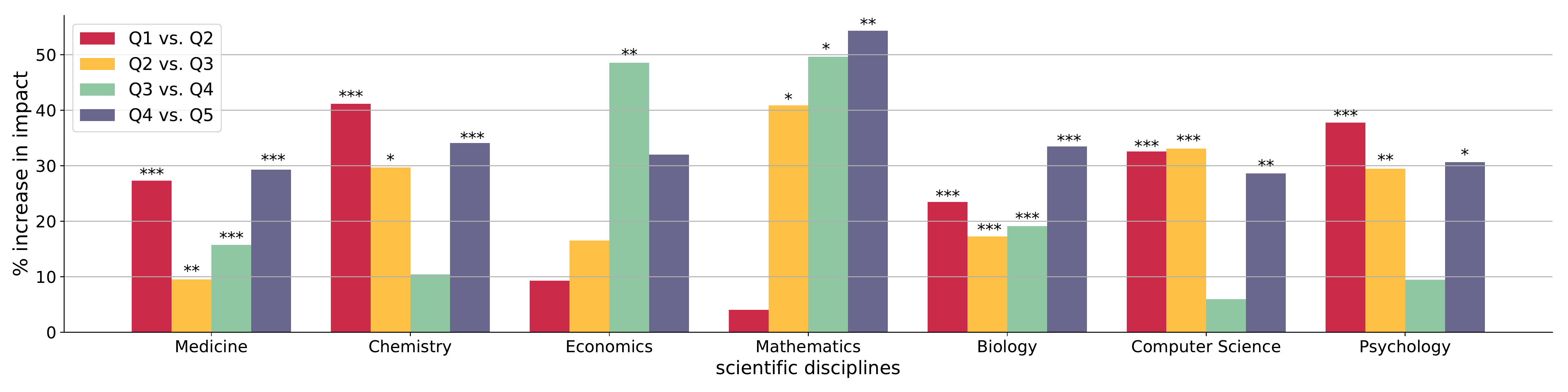}
\caption{\textbf{Big-shot effect across disciplines.} Here, $Q_i:i\in\{1,2,3,4\}$ represents the $i^{\textnormal{th}}$ quintile of the distribution of big-shot experience for all prot\'eg\'es who have the same discipline. Every bar represents a CEM in which the control and treatment groups are $Q_i$ and $Q_{i+1}$ for some $i\in\{1,2,3,4\}$, respectively (the color of the bar indicates the value of $i$). The bars are grouped based on disciplines. The height of the bar equals $\delta$, which is the increase in the post-mentorship impact of the treatment group relative to that of the control group. We omitted Geology and Engineering since they have the smallest numbers of matched pairs, and their $\delta$ values are all insignificant; see Supplementary Table~\ref{tab:CEMGroup:Discipline}. *p$<$0.05, **p$<$0.01, ***p$<$0.001.
}
\label{fig:}
\end{figure}

\begin{figure}[H]
\centering
\includegraphics[scale=.5]{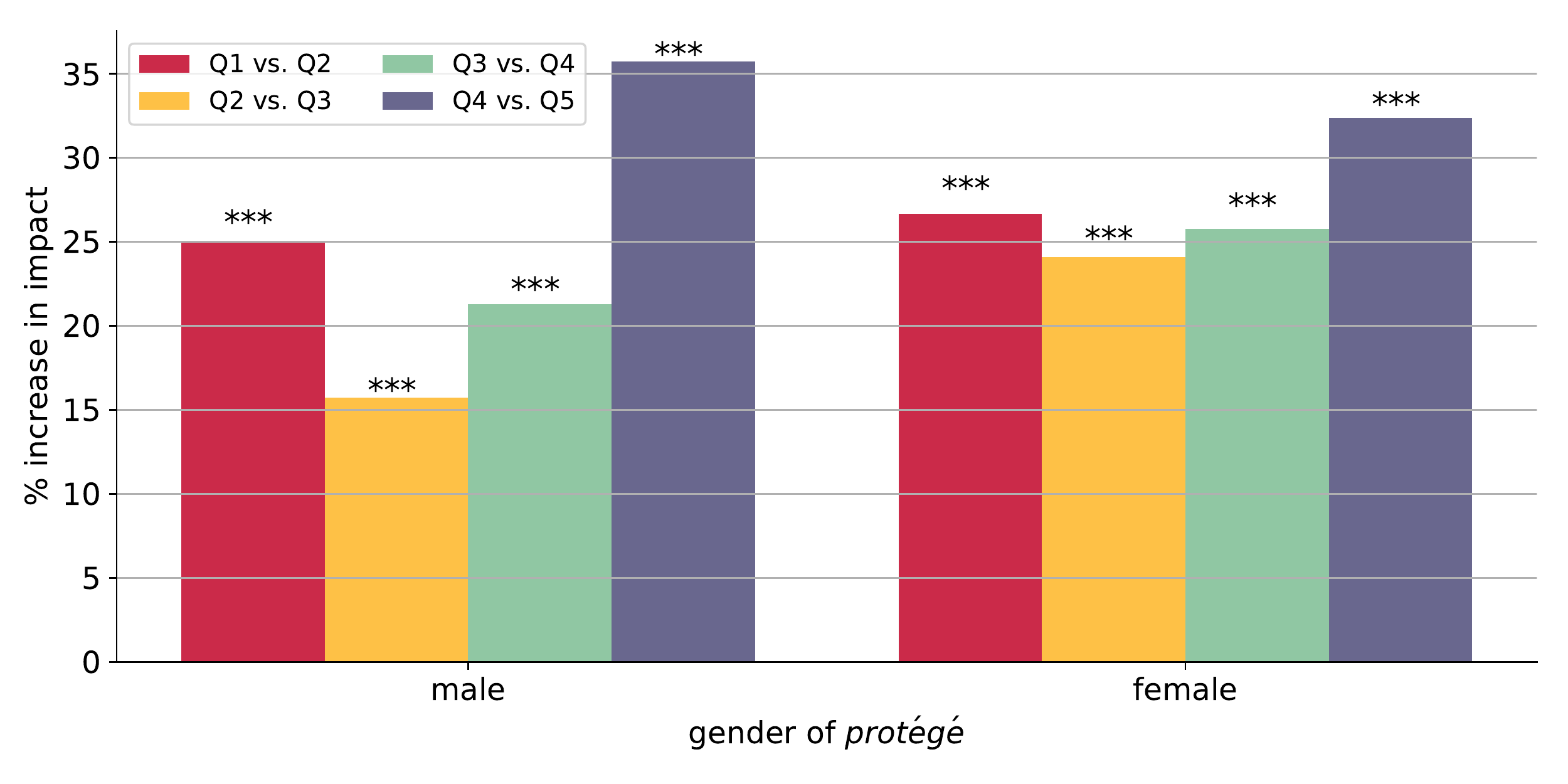}
\caption{\textbf{Big-shot effect between genders.} Here, $Q_i:i\in\{1,2,3,4\}$ represents the $i^{\textnormal{th}}$ quintile of the distribution of big-shot experience for all prot\'eg\'es who have the same gender. Every bar represents a CEM in which the control and treatment groups are $Q_i$ and $Q_{i+1}$ for some $i\in\{1,2,3,4\}$, respectively (the color of the bar indicates the value of $i$). The bars are grouped based on the gender of the prot\'eg\'e. The height of the bar equals $\delta$, which is the increase in the post-mentorship impact of the treatment group relative to that of the control group; see Supplementary Table~\ref{tab:CEMGroup:gender}. *p$<$0.05, **p$<$0.01, ***p$<$0.001.
}
\label{fig:}
\end{figure}

\begin{figure}[H]
\centering
\includegraphics[width=\linewidth]{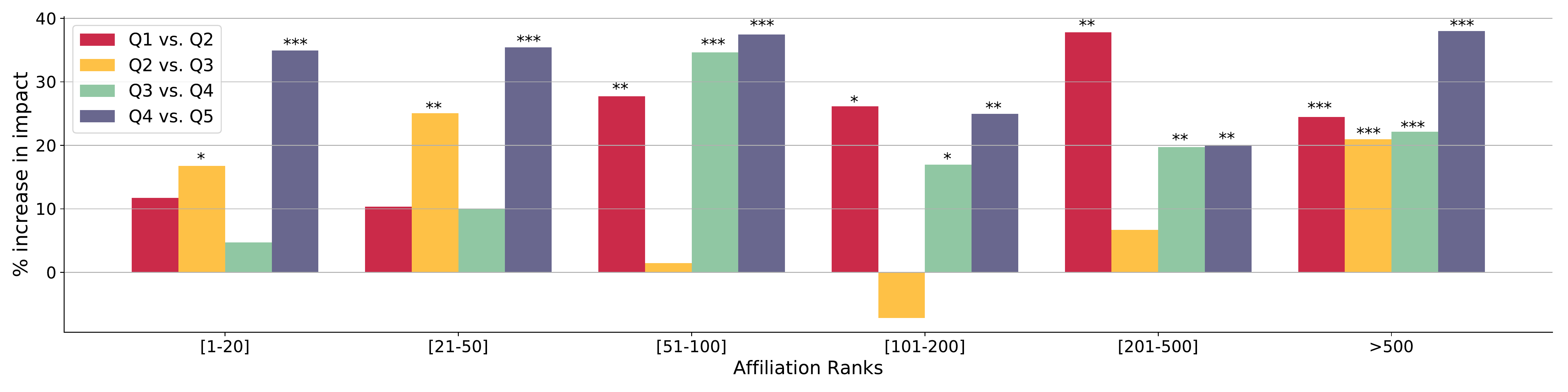}
\caption{\textbf{Big-shot effect across university ranks.} Here, $Q_i:i\in\{1,2,3,4\}$ represents the $i^{\textnormal{th}}$ quintile of the distribution of big-shot experience for all prot\'eg\'es whose affiliation on their first publication falls within the same bin of university ranks. Every bar represents a CEM in which the control and treatment groups are $Q_i$ and $Q_{i+1}$ for some $i\in\{1,2,3,4\}$, respectively (the color of the bar indicates the value of $i$). The bars are grouped based on the bins of the university ranks. The height of the bar equals $\delta$, which is the increase in the post-mentorship impact of the treatment group relative to that of the control group; see Supplementary Table~\ref{tab:CEMGroup:universityRank}. *p$<$0.05, **p$<$0.01, ***p$<$0.001.
}
\label{fig:}
\end{figure}

\clearpage

\section{Independent Variables and Confounding Factors in Coarsened Exact Matching}

This section outlines the variables used in the Coarsened Exact Matchings (CEMs), and specifies the binning decision for every such variable. The independent variables are indicated by ``IV'', whereas the confounding factors are indicated by ``C.''

\begin{enumerate}
    \item \textbf{Big-Shot experience (IV)}: This is computed for any given prot\'eg\'e by first computing the average annual number of citations of each mentor up to the year of their first publication with the prot\'eg\'e, and then averaging these numbers over all mentors. The data points are divided into 10 bins based on percentile cutoffs as follows: $<$4.0; 4.0-7.1; 7.2-10.5; 10.6-14.2; 14.3-18.4; 18.5-23.3, 23.4-29.5; 29.6-38.2; 38.3-54.2; $>$54.2.

    \item \textbf{Hub experience (IV)}: This is computed for any given prot\'eg\'e as the average degree of the mentors prior to mentorship, where the degree of each such mentor is measured in the network of scientific collaborations, i.e., the network where every node represents a scientist and an edge is added between two scientists if and only if they collaborated. Note that this network changes over time. The degree of each mentor is measured in the year of their first publication with the prot\'eg\'e. The data points are divided into 10 bins based on percentile cutoffs as follows: $<$18; 18-35; 36-57; 58-87; 88-131; 132-198; 199-310; 311- 525; 526-1109; $>$1109.
     
    \item \textbf{Number of mentors (C)}: The total number of mentors that the prot\'eg\'e has, which is divided into 5 bins based on quintile cutoffs as follows: 1; 2; 3; 4-5; $>$5.

    \item \textbf{Year of the prot\'eg\'e's first publication (C)}: The year in which the prot\'eg\'e published their first mentored paper. This is divided into 10 bins based on percentile cutoffs as follows: $<$1987; 1987-1992; 1993-1996; 1997-1998; 1999-2001; 2002-2003; 2004-2005; 2006-2007; 2008-2009; $\geq$ 2010.

 
    \item \textbf{Gender of the prot\'eg\'e (C)}: The data points in our study are divided into 2 bins: male and female. Following \cite{AlShebli_etal_2018},the gender of prot\'eg\'es and mentors is identified using a state-of-the art gender classifier \cite{Wais_2016}, and drop all mentor-prot\'eg\'e pairs where the gender of the prot\'eg\'e could not be established with at least 95\% certainty. Furthermore, when analyzing the gender of mentors, we also drop any mentors whose gender could not be established with at least 95\% certainty. 

    \item  \textbf{University rank (C)}: The \textit{rank}\footnote{University ranks are based on the 2018 \emph{``Academic Ranking of World Universities''}, also known as the \textit{``Shanghai ranking''}; see {\tt http://www.shanghairanking.com/ARWU2018.html}} of the affiliation of the prot\'eg\'e on their first mentored publication. The data points in our study are divided into the following bins: 1; 2; 3; \dots; 99; 100; 101-150; 151-200; 201-300; 301-400; 401-500; $>$500.

    \item \textbf{Average academic age of mentors (C)}: This is computed for any given prot\'eg\'e by first computing the \textit{academic age}\footnote{Given a scientist whose first paper was published in year $x$, the academic age of this scientist in year $y$ is $y-x$.} of each mentor in the year of their first publication with the prot\'eg\'e, and then averaging these numbers over all the mentors. The data points are divided into 10 bins based on percentile cutoffs as follows: 8-10.9; 11.0-13.3; 13.4-15.4; 15.5-17.2; 17.3-19.0; 19.1-21.0; 21.1-23.5; 23.6-26.7; 26.8-32.0; $\geq$ 32.1.
        
        
     \item \textbf{The number of years post mentorship (C)}: Since our dataset was obtained in October 2015, we are only able to calculate $c_5$---the number of citations accumulated five years post publication---for papers published before 2011. Thus, given a prot\'eg\'e whose first paper was published in year $x$, the number of years post mentorship is $2011 - x - 7$, bearing in mind that the mentorship period is 7 years. 
     The data points in our study are divided into 10 bins based on percentile cutoffs as follows: 1; 2; 3; 4; 5-6; 7-8; 9-11; 12-14; 15-19; $\geq20$.
     
    \item \textbf{Scientific discipline (C)}: The prot\'eg\'es are classified into disciplines using the method proposed by \cite{AlShebli_etal_2018}. This method classifies any given scientist, $s_i$, based on the keywords that are specified in the publications of $s_i$, which are themselves classified by Microsoft Academic Graph (MAG) into disciplines; more details can be found in \cite{AlShebli_etal_2018}. We focus on the 10 disciplines in MAG that have the largest number of mentor-prot\'eg\'e pairs; see Supplementary Table~\ref{tab:num_pairs_discipline}. Out of these 10 disciplines, we exclude Physics since the mean number of authors per paper in this discipline is far larger than that of any other discipline (236, compared to 13 or below; see Supplementary Table~\ref{tab:mean_groupsize_discipline}). As such, keeping Physics would render our approach of identifying mentor-prot\'eg\'e relationships quixotic and impractical. Consequently, the data points in our study are divided into the following 9 disciplines: Biology, Computer Science, Chemistry, Economics, Engineering, Geology, Mathematics, Medicine, and Psychology.

    \end{enumerate}            

\begin{table}[htbp]
\centering
{\fontsize{12}{12}\selectfont{
\begin{tabular}{lr}
    \toprule
    Discipline & Number of mentor-prot\'eg\'e\\
    & pairs identified\\
    \midrule
    \textbf{Biology} & 234571 \\   
    \textbf{Physics} & 177582 \\              
    \textbf{Medicine} & 148079 \\            
    \textbf{Chemistry} & 101523 \\      
    \textbf{Computer Science} & 82759 \\           
    \textbf{Psychology} & 46544 \\           
    \textbf{Mathematics} & 34986 \\        
    \textbf{Engineering} & 17440 \\             
    \textbf{Economics} & 14254 \\               
    \textbf{Geology} & 12381 \\   
    Materials Science & 11175 \\            
    Sociology & 6710 \\           
    Philosophy & 995 \\             
    Business & 462 \\ 
    Geography & 306 \\            
    History & 228 \\ 
    Environmental science &184 \\    
    Political Science & 150 \\           
    Art & 150 \\
    \bottomrule
\end{tabular}
}}
\caption{The scientific disciplines in MAG, sorted based on the number of mentor-prot\'eg\'e pairs identified in each discipline. The bold font highlights the 10 disciplines containing the largest number of such pairs.}
\label{tab:num_pairs_discipline}
\end{table}

\begin{table}[htbp] 
\centering
{\fontsize{12}{12}\selectfont{
\begin{tabular}{lrr}
\toprule               
    Discipline &  95th quantile &  Mean number of \\
    & & authors per paper\\ 
    \midrule
    Economics &            7.0 &        3.82 \\
    Engineering &           11.0 &        5.10 \\
    Computer Science &           13.0 &        5.64 \\
    Mathematics &            9.0 &        6.23 \\
    Psychology &           12.0 &        6.35 \\
    Chemistry &           13.0 &        6.64 \\
    Geology &           17.0 &        7.45 \\
    Medicine &           16.0 &        8.67 \\
    Biology &           24.0 &       13.35 \\
    \textbf{Physics} & \textbf{2973.0} & \textbf{235.91} \\
\bottomrule
\end{tabular}
}}
\caption{For each of the 10 disciplines that are highlighted in Supplementary Table~\ref{tab:num_pairs_discipline}, the table specifies the $95^{\textnormal{th}}$ quantile and the mean number of authors per paper. The 10 disciplines are sorted based on mean number of authors per paper in each discipline. As can seen, Physics is an outlier, which is why it is excluded from our analyses.}
\label{tab:mean_groupsize_discipline}
\end{table}

\clearpage

\section{Results of Coarsened Exact Matching}

In this section, we present the results of all Coarsened Exact Matchings (CEMs). Before presenting these results, we need to introduce the following notation:
\begin{itemize}
\item $C$ and $T$ denote the control and treatment groups of prot\'eg\'es, respectively. In more detail, let $Q_i$ be the $i^{\textnormal{th}}$ quintile of the distribution of the independent variable under consideration, be it the big-shot experience or the hub experience. Then, For $i\in\{1,2,3,4\}$, we build a separate CEM where $C = Q_i$ and $T = Q_{i+1}$.
\item $C'$ and $T'$ denote the \textit{matched} control and treatment groups, respectively.
\item $\mathit{imp}(C')$ and $\mathit{imp}(T')$ denote the mean post-mentorship impact of the prot\'eg\'es in $C'$ and $T'$, respectively.
item $\mathcal{L}_1$ denotes the multivariate imbalance statistic in any given CEM; for mor details, see~\cite{iacus2012causal};
\item $\delta$ represents the relative difference in impact between the prot\'eg\'es in $C'$ and $T'$. Formally:
$$
\delta = 100\cdot \frac{\mathit{imp}(T') - \mathit{imp}(C')}{\mathit{imp}(C')}
$$
\item $p$: We conduct a t-test to determine the statistical significance between $\mathit{imp}(C')$ and $\mathit{imp}(T')$, and $p$ indicates the corresponding p-values of the test. Note that the Kolmogorov-Smirnov test was also applied, and similar statistical significance was observed.
\end{itemize}
With this notation in place, we are now ready to present the results of the different CEMs.

\vspace{0.7cm}\hspace{-0.7cm}
\begin{minipage}{\textwidth}

\begin{table}[H]
{\fontsize{10}{10}\selectfont{
\caption{Results of the CEMs that were used to quantify the effect of the \textbf{big-shot experience} on the scientific impact of the prot\'eg\'e post mentorship. Here, for $i\in\{1,2,3,4\}$, we write $Q_i$ vs.~$Q_{i+1}$ to indicate that $C=Q_i$ and $T=Q_{i+1}$.}
\label{tab:CEMGroup:bigshot}
\begin{center}
\begin{tabular}{lcccccccccccc}

\toprule
 & $|T|$ & $|C|$ &  $|T'|$ & $|C'|$ & $\mathcal{L}_1$ & $\mathit{imp}(C')$ & $\mathit{imp}(T')$ & $\delta$(\%) & $p$\\
\midrule
$Q_1$ vs.~$Q_2$ & 94310 & 94310 & 19988 & 24150 & 
0.3 & 14.86 & 11.66 & 
27.4 & 4.0e-36\\

\midrule
$Q_2$ vs.~$Q_3$ & 94309 & 94311 & 21050 & 20760 & 
0.25 & 16.35 & 14.05 & 
16.4 & 1.4e-17\\

\midrule
$Q_3$ vs.~$Q_4$ & 94309 & 94311 & 22356 & 22105 & 
0.26 & 19.47 & 16.03 & 
21.5 & 1.2e-30\\

\midrule
$Q_4$ vs.~$Q_5$ & 94308 & 94311 & 24654 & 22415 & 
0.28 & 28.26 & 20.86 & 
35.5 & 2.8e-77\\

\midrule
\end{tabular}
\end{center}
}}
\end{table}

\end{minipage}

\vspace{0.7cm}\hspace{-0.7cm}
\begin{minipage}{\textwidth}

\begin{table}[H]
{\fontsize{10}{10}\selectfont{
\caption{Results of the CEMs that were used to quantify the effect of the \textbf{hub experience} on the scientific impact of the prot\'eg\'e post mentorship. Here, for $i\in\{1,2,3,4\}$, we write $Q_i$ vs.~$Q_{i+1}$ to indicate that $C=Q_i$ and $T=Q_{i+1}$.}
\label{tab:CEMGroup:hub}
\begin{center}
\begin{tabular}{lcccccccccccc}

\toprule
 & $|T|$ & $|C|$ &  $|T'|$ & $|C'|$ & $\mathcal{L}_1$ & $\mathit{imp}(C')$ & $\mathit{imp}(T')$ & $\delta$(\%) & $p$\\
\midrule
$Q_1$ vs.~$Q_2$ & 93962 & 95775 & 18374 & 23499 & 
0.31 & 19.83 & 18.61 & 
6.6 & 0.0007 \\

\midrule
$Q_2$ vs.~$Q_3$ & 93397 & 96544 & 16208 & 17963 & 
0.27 & 20.83 & 19.97 & 
4.4 & 0.03\\

\midrule
$Q_3$ vs.~$Q_4$ & 94165 & 94791 & 17596 & 16875 & 
0.26 & 19.91 & 19.1 & 
4.2 & 0.04\\

\midrule
$Q_4$ vs.~$Q_5$ & 94247 & 94690 & 25099 & 19013 & 
0.35 & 16.98 & 15.86 & 
7.1 & 0.001\\

\midrule
\end{tabular}
\end{center}
}}
\end{table}

\end{minipage}

\vspace{0.7cm}\hspace{-0.7cm}
\begin{minipage}{\textwidth}

\begin{table}[H]
{\fontsize{10}{10}\selectfont{
\caption{Results of the CEMs used to quantify the effect of the big-shot experience on the scientific impact of the prot\'eg\'e post mentorship while \textbf{controlling for the year of the prot\'eg\'e's first publication}. For $i\in\{1,2,3,4\}$, we write $Q_i$ vs.~$Q_{i+1}$ to indicate that $C=Q_i$ and $T=Q_{i+1}$.}
\label{tab:CEMGroup:yearOfFirstPublication}
\begin{center}
\begin{tabular}{llcccccccccccc}

\toprule
Year && $|C|$ & $|T|$ &  $|C'|$ & $|T'|$ & $\mathcal{L}_1$ & $\mathit{imp}(C')$ & $\mathit{imp}(T')$ & $\delta$(\%) & $p$\\
\midrule
$<'90$ & $Q_1$ vs.~$Q_2$ & 10836 & 10837 & 4238 & 4976 & 
0.29 & 23.11 & 19.34 & 
19.5 & 3.9e-05\\ 
& $Q_2$ vs.~$Q_3$ & 10838 & 10836 & 3912 & 4290 & 
0.25 & 28.82 & 24.97 & 
15.4 & 8.6e-07\\ 
& $Q_3$ vs.~$Q_4$ & 10835 & 10838 & 3890 & 3986 & 
0.23 & 34.16 & 30.72 & 
11.2 & 0.0003\\ 
& $Q_4$ vs.~$Q_5$ & 10837 & 10837 & 3898 & 3896 & 
0.23 & 45.36 & 35.82 & 
26.6 & 5.9e-15\\ 
\midrule
$['90,'95)$ & $Q_1$ vs.~$Q_2$ & 9023 & 9023 & 1911 & 2392 & 0.31 & 19.11 & 16.79 & 13.9 & 0.046\\ 
& $Q_2$ vs.~$Q_3$ & 9022 & 9023 & 1877 & 1914 & 
0.25 & 26.01 & 23.03 & 
12.9 & 0.04\\ 
& $Q_3$ vs.~$Q_4$ & 9024 & 9022 & 1873 & 1859 & 
0.26 & 36.11 & 28.77 & 
25.5 & 5.9e-08\\ 
& $Q_4$ vs.~$Q_5$ & 9022 & 9024 & 2070 & 1941 & 
0.27 & 47.99 & 37.82 & 
26.9 & 3.5e-08\\ 
\midrule
$['95,'00)$ & $Q_1$ vs.~$Q_2$ & 17100 & 17101 & 4259 & 4891 & 0.3 & 18.43 & 14.51 & 27.0 & 6.9e-11\\ 
& $Q_2$ vs.~$Q_3$ & 17100 & 17101 & 4652 & 4605 & 
0.26 & 23.62 & 20.38 & 15.9 & 7.8e-07\\ 
& $Q_3$ vs.~$Q_4$ & 17102 & 17101 & 4790 & 4819 & 
0.27 & 31.07 & 24.18 & 28.5 & 2.8e-20\\ 
& $Q_4$ vs.~$Q_5$ & 17098 & 17104 & 5490 & 4725 & 
0.29 & 41.81 & 31.58 & 
32.4 & 4.95e-26\\ 
\midrule
$['00,'05)$ & $Q_1$ vs.~$Q_2$ & 24244 & 24245 & 6118 & 6870 & 0.3 & 11.7 & 8.66 & 35.1 & 5.7e-17\\ 
& $Q_2$ vs.~$Q_3$ & 24245 & 24244 & 6960 & 6571 & 
0.28 & 15.42 & 12.37 & 24.6 & 6.6e-16\\ 
& $Q_3$ vs.~$Q_4$ & 24243 & 24245 & 7679 & 7257 & 
0.28 & 19.76 & 16.61 & 19.0 & 1.1e-09\\ 
& $Q_4$ vs.~$Q_5$ & 24245 & 24247 & 8456 & 7626 & 
0.3 & 28.9 & 20.57 & 40.5 & 8.9e-41\\ 
\midrule
$['05,'10)$ & $Q_1$ vs.~$Q_2$ & 25377 & 25380 & 7437 & 8183 & 0.34 & 4.94 & 3.56 & 38.6 & 8.1e-13\\ 
& $Q_2$ vs.~$Q_3$ & 25380 & 25379 & 8513 & 7800 & 
0.28 & 5.86 & 4.96 & 18.3 & 2.9e-08\\ 
& $Q_3$ vs.~$Q_4$ & 25377 & 25380 & 9369 & 8750 & 
0.27 & 8.52 & 6.63 & 28.5 & 1.0e-19\\ 
& $Q_4$ vs.~$Q_5$ & 25379 & 25379 & 9817 & 9066 & 
0.29 & 14.28 & 9.79 & 45.9 & 9.9e-28\\ 
\midrule
\end{tabular}
\end{center}
}}
\end{table}

\end{minipage}

\vspace{0.7cm}\hspace{-0.7cm}
\begin{minipage}{\textwidth}

\begin{table}[H]
{\fontsize{10}{10}\selectfont{
\caption{Results of the CEMs used to quantify the effect of the big-shot experience on the scientific impact of the prot\'eg\'e post mentorship while \textbf{controlling for the average academic age of mentors}. For $i\in\{1,2,3,4\}$, we write $Q_i$ vs.~$Q_{i+1}$ to indicate that $C=Q_i$ and $T=Q_{i+1}$.}
\label{tab:CEMGroup:AgeOfMentors}
\begin{center}
\begin{tabular}{llcccccccccccc}

\toprule
Age && $|C|$ & $|T|$ &  $|C'|$ & $|T'|$ & $\mathcal{L}_1$ & $\mathit{imp}(C')$ & $\mathit{imp}(T')$ & $\delta$(\%) & $p$\\
\midrule
$[8,11)$ & $Q_1$ vs.~$Q_2$ & 9138 & 9138 & 3547 & 4232 & 
0.29 & 14.28 & 12.01 & 18.9 & 0.0001\\ 
& $Q_2$ vs.~$Q_3$ & 9138 & 9138 & 3273 & 3449 & 
0.24 & 17.77 & 16.11 & 10.3 & 0.01\\ 
& $Q_3$ vs.~$Q_4$ & 9135 & 9141 & 3014 & 3232 & 
0.24 & 22.7 & 19.13 & 18.7 & 0.0006\\ 
& $Q_4$ vs.~$Q_5$ & 9138 & 9138 & 2756 & 2824 & 
0.25 & 28.89 & 22.34 & 29.3 & 7.8e-09\\ 
\midrule
$[11,15)$ & $Q_1$ vs.~$Q_2$ & 17103 & 17104 & 5665 & 6813 & 0.3 & 15.35 & 12.23 & 25.6 & 3.8e-08\\ 
& $Q_2$ vs.~$Q_3$ & 17104 & 17104 & 5405 & 5546 & 
0.25 & 19.18 & 16.73 & 14.7 & 9.4e-06\\ 
& $Q_3$ vs.~$Q_4$ & 17102 & 17105 & 5234 & 5268 & 
0.25 & 23.88 & 19.64 & 21.6 & 4.5e-11\\ 
& $Q_4$ vs.~$Q_5$ & 17103 & 17106 & 5369 & 5249 & 
0.28 & 30.33 & 22.72 & 33.5 & 1.2e-24\\ 
\midrule
$[15,20)$ & $Q_1$ vs.~$Q_2$ & 25895 & 25898 & 8448 & 9020 & 0.32 & 13.24 & 10.38 & 27.6 & 1.6e-14\\ 
& $Q_2$ vs.~$Q_3$ & 25897 & 25897 & 9318 & 8926 & 
0.28 & 15.76 & 13.72 & 14.9 & 5.5e-08\\ 
& $Q_3$ vs.~$Q_4$ & 25896 & 25897 & 9755 & 9585 & 
0.28 & 20.19 & 16.24 & 24.4 & 3.9e-21\\ 
& $Q_4$ vs.~$Q_5$ & 25897 & 25896 & 10410 & 9461 & 
0.29 & 28.82 & 21.36 & 34.9 & 5.4e-40\\ 
\midrule
$[20,25)$ & $Q_1$ vs.~$Q_2$ & 19643 & 19644 & 5205 & 5072 & 0.31 & 12.16 & 9.18 & 32.4 & 4.0e-12\\ 
& $Q_2$ vs.~$Q_3$ & 19643 & 19644 & 6457 & 5803 & 
0.28 & 14.36 & 11.89 & 20.7 & 5.1e-07\\ 
& $Q_3$ vs.~$Q_4$ & 19643 & 19645 & 6937 & 6663 & 
0.27 & 18.36 & 14.27 & 28.6 & 9.0e-17\\ 
& $Q_4$ vs.~$Q_5$ & 19643 & 19644 & 7452 & 6728 & 
0.3 & 27.99 & 20.18 & 38.7 & 1.6e-29\\ 
\midrule
$[25,30)$ & $Q_1$ vs.~$Q_2$ & 10627 & 10627 & 1742 & 1740 & 0.27 & 11.02 & 9.09 & 21.2 & 0.002\\ 
& $Q_2$ vs.~$Q_3$ & 10626 & 10627 & 2325 & 2027 & 
0.28 & 12.84 & 9.94 & 29.2 & 4.9e-05\\ 
& $Q_3$ vs.~$Q_4$ & 10626 & 10629 & 2595 & 2499 & 
0.27 & 16.25 & 12.72 & 27.7 & 1.4e-05\\ 
& $Q_4$ vs.~$Q_5$ & 10627 & 10626 & 3019 & 2613 & 
0.29 & 27.56 & 18.81 & 46.5 & 8.4e-16\\ 
\midrule
$\geq 30$ & $Q_1$ vs.~$Q_2$ & 11902 & 11903 & 2101 & 2252 & 0.29 & 12.41 & 10.2 & 21.6 & 0.001\\ 
& $Q_2$ vs.~$Q_3$ & 11902 & 11903 & 2400 & 2315 & 
0.25 & 12.69 & 11.28 & 12.5 & 0.04\\ 
& $Q_3$ vs.~$Q_4$ & 11903 & 11903 & 2586 & 2589 & 
0.24 & 16.37 & 13.3 & 23.1 & 1.6e-05\\ 
& $Q_4$ vs.~$Q_5$ & 11901 & 11906 & 2985 & 2643 & 
0.28 & 25.44 & 17.8 & 42.9 & 1.8e-15\\ 
\midrule
\end{tabular}
\end{center}
}}
\end{table}

\end{minipage}

\vspace{0.7cm}\hspace{-0.7cm}
\begin{minipage}{\textwidth}

\begin{table}[H]
{\fontsize{10}{10}\selectfont{
\caption{Results of the CEMs used to quantify the effect of the big-shot experience on the scientific impact of the prot\'eg\'e post mentorship while \textbf{controlling for the number of mentors}. For $i\in\{1,2,3,4\}$, we write $Q_i$ vs.~$Q_{i+1}$ to indicate that $C=Q_i$ and $T=Q_{i+1}$.}
\label{tab:CEMGroup:NumberOfAuthors}
\begin{center}
\begin{tabular}{llcccccccccccc}

\toprule
No.~mentors && $|C|$ & $|T|$ &  $|C'|$ & $|T'|$ & $\mathcal{L}_1$ & $\mathit{imp}(C')$ & $\mathit{imp}(T')$ & $\delta$(\%) & $p$\\
\midrule
$1$ & $Q_1$ vs.~$Q_2$ & 33923 & 33923 & 9753 & 11621 & 0.29 & 14.85 & 12.53 & 18.5 & 1.8e-07\\ 
& $Q_2$ vs.~$Q_3$ & 33926 & 33923 & 8906 & 9393 & 
0.24 & 18.47 & 16.09 & 14.8 & 4.6e-08\\ 
& $Q_3$ vs.~$Q_4$ & 33919 & 33926 & 8621 & 8923 & 
0.24 & 21.43 & 18.57 & 15.4 & 5.3e-09\\ 
& $Q_4$ vs.~$Q_5$ & 33923 & 33925 & 8544 & 8413 & 
0.24 & 28.76 & 22.1 & 30.1 & 10.0e-25\\ 
\midrule
$2$ & $Q_1$ vs.~$Q_2$ & 12449 & 12447 & 1708 & 1657 & 0.24 & 13.82 & 11.21 & 23.3 & 0.002\\ 
& $Q_2$ vs.~$Q_3$ & 12445 & 12449 & 1883 & 1840 & 0.24 & 16.84 & 14.67 & 14.8 & 0.01\\ 
& $Q_3$ vs.~$Q_4$ & 12447 & 12448 & 2046 & 2020 & 
0.24 & 20.29 & 16.51 & 22.9 & 2.1e-05\\ 
& $Q_4$ vs.~$Q_5$ & 12446 & 12450 & 2348 & 2084 & 
0.27 & 28.32 & 21.96 & 29.0 & 3.8e-08\\ 
\midrule
$3$ & $Q_1$ vs.~$Q_2$ & 12449 & 12447 & 1708 & 1657 & 
0.24 & 13.82 & 11.21 & 23.3 & 0.002\\ 
& $Q_2$ vs.~$Q_3$ & 12445 & 12449 & 1883 & 1840 & 
0.24 & 16.84 & 14.67 & 14.8 & 0.01\\ 
& $Q_3$ vs.~$Q_4$ & 12447 & 12448 & 2046 & 2020 & 
0.24 & 20.29 & 16.51 & 22.9 & 2.1e-05\\ 
& $Q_4$ vs.~$Q_5$ & 12446 & 12450 & 2348 & 2084 & 
0.27 & 28.32 & 21.96 & 29.0 & 3.8e-08\\ 
\midrule
$[4,5]$ & $Q_1$ vs.~$Q_2$ & 13588 & 13589 & 2395 & 2260 & 0.26 & 12.02 & 9.26 & 29.8 & 3.9e-06\\ 
& $Q_2$ vs.~$Q_3$ & 13588 & 13588 & 2592 & 2625 & 
0.26 & 13.99 & 12.19 & 14.8 & 0.006\\ 
& $Q_3$ vs.~$Q_4$ & 13589 & 13588 & 2855 & 2765 & 
0.27 & 18.73 & 14.74 & 27.1 & 6.7e-07\\ 
& $Q_4$ vs.~$Q_5$ & 13589 & 13589 & 3428 & 3066 & 
0.29 & 27.52 & 19.88 & 38.4 & 1.0e-16\\ 
\midrule
$>5$ & $Q_1$ vs.~$Q_2$ & 14529 & 14535 & 4376 & 3784 & 
0.31 & 9.18 & 6.6 & 39.0 & 5.6e-16\\ 
& $Q_2$ vs.~$Q_3$ & 14533 & 14533 & 4980 & 4919 & 
0.3 & 13.24 & 10.38 & 27.6 & 2.1e-10\\ 
& $Q_3$ vs.~$Q_4$ & 14531 & 14533 & 4970 & 5026 & 
0.29 & 17.08 & 14.18 & 20.5 & 2.0e-08\\ 
& $Q_4$ vs.~$Q_5$ & 14532 & 14534 & 5859 & 4761 & 
0.31 & 28.7 & 20.38 & 40.8 & 1.8e-16\\ 
\midrule
\end{tabular}
\end{center}
}}
\end{table}

\end{minipage}

\vspace{0.7cm}\hspace{-0.7cm}
\begin{minipage}{\textwidth}

\begin{table}[H]
{\fontsize{10}{10}\selectfont{
\caption{Results of the CEMs used to quantify the effect of the big-shot experience on the scientific impact of the prot\'eg\'e post mentorship while \textbf{controlling for university rank}. For $i\in\{1,2,3,4\}$, we write $Q_i$ vs.~$Q_{i+1}$ to indicate that $C=Q_i$ and $T=Q_{i+1}$.}
\label{tab:CEMGroup:universityRank}
\begin{center}
\begin{tabular}{llcccccccccccc}

\toprule
University rank && $|C|$ & $|T|$ &  $|C'|$ & $|T'|$ & $\mathcal{L}_1$ & $\mathit{imp}(C')$ & $\mathit{imp}(T')$ & $\delta$(\%) & $p$\\
\midrule
$[1,20]$ & $Q_1$ vs.~$Q_2$ & 10050 & 10049 & 1320 & 1613 & 0.27 & 23.85 & 21.34 & 11.7 & 0.1\\ 
& $Q_2$ vs.~$Q_3$ & 10046 & 10051 & 1434 & 1464 & 
0.2 & 23.95 & 20.51 & 16.8 & 0.02\\ 
& $Q_3$ vs.~$Q_4$ & 10048 & 10049 & 1520 & 1477 & 
0.21 & 22.91 & 21.87 & 4.7 & 0.5\\ 
& $Q_4$ vs.~$Q_5$ & 10048 & 10049 & 1825 & 1714 & 
0.21 & 34.18 & 25.33 & 35.0 & 2.3e-09\\ 
\midrule
$[21,50]$ & $Q_1$ vs.~$Q_2$ & 7040 & 7043 & 774 & 1006 & 0.27 & 21.43 & 19.41 & 10.4 & 0.2\\ 
& $Q_2$ vs.~$Q_3$ & 7041 & 7042 & 809 & 807 & 
0.18 & 19.59 & 15.67 & 25.1 & 0.003\\ 
& $Q_3$ vs.~$Q_4$ & 7041 & 7043 & 875 & 918 & 
0.19 & 22.36 & 20.33 & 10.0 & 0.2\\ 
& $Q_4$ vs.~$Q_5$ & 7041 & 7042 & 966 & 926 & 
0.19 & 30.31 & 22.38 & 35.4 & 5.7e-06\\ 
\midrule
$[51,100]$ & $Q_1$ vs.~$Q_2$ & 8026 & 8026 & 883 & 1082 & 0.23 & 16.65 & 13.04 & 27.7 & 0.002\\ 
& $Q_2$ vs.~$Q_3$ & 8026 & 8026 & 834 & 847 & 
0.17 & 14.83 & 14.61 & 1.5 & 0.8\\ 
& $Q_3$ vs.~$Q_4$ & 8026 & 8026 & 986 & 920 & 
0.18 & 16.74 & 12.43 & 34.7 & 0.0003\\ 
& $Q_4$ vs.~$Q_5$ & 8026 & 8026 & 1083 & 1084 & 
0.18 & 21.44 & 15.6 & 37.5 & 3.5e-07\\ 
\midrule
$[101,200]$ & $Q_1$ vs.~$Q_2$ & 6028 & 6044 & 609 & 817 & 0.29 & 20.47 & 16.23 & 26.1 & 0.03\\ 
& $Q_2$ vs.~$Q_3$ & 6035 & 6039 & 538 & 589 & 
0.19 & 17.4 & 18.75 & -7.2 & 0.5\\ 
& $Q_3$ vs.~$Q_4$ & 6036 & 6035 & 688 & 679 & 
0.19 & 15.77 & 13.48 & 17.0 & 0.03\\ 
& $Q_4$ vs.~$Q_5$ & 6036 & 6036 & 743 & 711 & 
0.18 & 22.72 & 18.18 & 25.0 & 0.004\\ 
\midrule
$[201,500]$ & $Q_1$ vs.~$Q_2$ & 9201 & 9200 & 1142 & 1480 & 0.26 & 16.45 & 11.94 & 37.8 & 0.004\\ 
& $Q_2$ vs.~$Q_3$ & 9197 & 9201 & 967 & 1060 & 
0.2 & 15.92 & 14.93 & 6.7 & 0.4\\ 
& $Q_3$ vs.~$Q_4$ & 9199 & 9200 & 1079 & 1064 & 
0.17 & 16.52 & 13.8 & 19.7 & 0.007\\ 
& $Q_4$ vs.~$Q_5$ & 9200 & 9199 & 1325 & 1349 & 
0.19 & 20.99 & 17.5 & 19.9 & 0.002\\ 
\midrule
$>500$ & $Q_1$ vs.~$Q_2$ & 53904 & 54013 & 16809 & 19827 & 0.3 & 13.84 & 11.11 & 24.5 & 1.6e-24\\ 
& $Q_2$ vs.~$Q_3$ & 53959 & 53964 & 19302 & 18692 & 0.27 & 15.73 & 13.01 & 21.0 & 3.1e-25\\ 
& $Q_3$ vs.~$Q_4$ & 53959 & 53959 & 21923 & 20992 & 0.28 & 18.97 & 15.53 & 22.1 & 1.7e-31\\ 
& $Q_4$ vs.~$Q_5$ & 53958 & 53959 & 24978 & 22602 & 0.3 & 27.93 & 20.24 & 38.0 & 1.0e-88\\ 
\midrule
\end{tabular}
\end{center}
}}
\end{table}

\end{minipage}

\vspace{0.7cm}\hspace{-0.7cm}
\begin{minipage}{\textwidth}

\begin{table}[H]
{\fontsize{10}{10}\selectfont{
\caption{Results of the CEMs used to quantify the effect of the big-shot experience on the scientific impact of the prot\'eg\'e post mentorship while \textbf{controlling for the gender of the prot\'eg\'e}. For $i\in\{1,2,3,4\}$, we write $Q_i$ vs.~$Q_{i+1}$ to indicate that $C=Q_i$ and $T=Q_{i+1}$.}
\label{tab:CEMGroup:gender}
\begin{center}
\begin{tabular}{llcccccccccccc}

\toprule
Gender && $|C|$ & $|T|$ &  $|C'|$ & $|T'|$ & $\mathcal{L}_1$ & $\mathit{imp}(C')$ & $\mathit{imp}(T')$ & $\delta$(\%) & $p$\\
\midrule
Female & $Q_1$ vs.~$Q_2$ & 32198 & 32180 & 5911 & 6499 & 0.28 & 12.12 & 9.57 & 26.7 & 1.6e-10\\ 
& $Q_2$ vs.~$Q_3$ & 32162 & 32198 & 6485 & 6419 & 
0.24 & 13.14 & 10.59 & 24.1 & 6.0e-11\\ 
& $Q_3$ vs.~$Q_4$ & 32180 & 32189 & 6936 & 6853 & 
0.24 & 16.34 & 12.99 & 25.8 & 7.7e-13\\ 
& $Q_4$ vs.~$Q_5$ & 32180 & 32180 & 7736 & 7079 & 
0.28 & 24.37 & 18.41 & 32.4 & 1.4e-16\\ 
\midrule
Male & $Q_1$ vs.~$Q_2$ & 57745 & 57781 & 14001 & 17302 & 0.3 & 15.36 & 12.29 & 25.0 & 9.2e-22\\ 
& $Q_2$ vs.~$Q_3$ & 57763 & 57852 & 14260 & 14302 & 0.26 & 17.7 & 15.29 & 15.7 & 3.5e-11\\ 
& $Q_3$ vs.~$Q_4$ & 57763 & 57763 & 15166 & 14784 & 0.27 & 21.32 & 17.58 & 21.3 & 9.8e-24\\ 
& $Q_4$ vs.~$Q_5$ & 57763 & 57763 & 16736 & 15225 & 0.29 & 30.33 & 22.34 & 35.7 & 1.2e-61\\ 
\midrule
\end{tabular}
\end{center}
}}
\end{table}

\end{minipage}

\vspace{0.7cm}\hspace{-0.7cm}
\begin{minipage}{\textwidth}

\begin{table}[H]
{\fontsize{10}{10}\selectfont{
\caption{Results of the CEMs used to quantify the effect of the big-shot experience on the scientific impact of the prot\'eg\'e post mentorship while \textbf{controlling for discipline}. For $i\in\{1,2,3,4\}$, we write $Q_i$ vs.~$Q_{i+1}$ to indicate that $C=Q_i$ and $T=Q_{i+1}$.}
\label{tab:CEMGroup:Discipline}
\begin{center}
\begin{tabular}{llcccccccccccc}

\toprule
Discipline && $|C|$ & $|T|$ &  $|C'|$ & $|T'|$ & $\mathcal{L}_1$ & $\mathit{imp}(C')$ & $\mathit{imp}(T')$ & $\delta$(\%) & $p$\\
\midrule
Medicine & $Q_1$ vs.~$Q_2$ & 22505 & 22509 & 6187 & 7733 & 0.31 & 14.17 & 11.13 & 27.3 & 1.1e-14\\ 
& $Q_2$ vs.~$Q_3$ & 22506 & 22508 & 6069 & 6120 & 
0.25 & 13.97 & 12.75 & 9.6 & 0.005\\ 
& $Q_3$ vs.~$Q_4$ & 22506 & 22508 & 6040 & 6021 & 
0.26 & 16.97 & 14.66 & 15.7 & 7.8e-06\\ 
& $Q_4$ vs.~$Q_5$ & 22507 & 22506 & 5573 & 5758 & 
0.26 & 24.56 & 18.99 & 29.3 & 5.6e-18\\ 
\midrule
Chemistry & $Q_1$ vs.~$Q_2$ & 10244 & 10257 & 1881 & 2360 & 0.25 & 10.77 & 7.63 & 41.2 & 3.1e-10\\ 
& $Q_2$ vs.~$Q_3$ & 10250 & 10254 & 1698 & 1780 & 
0.19 & 12.57 & 9.69 & 29.7 & 0.01\\ 
& $Q_3$ vs.~$Q_4$ & 10250 & 10251 & 1380 & 1448 & 
0.18 & 12.15 & 11.0 & 10.4 & 0.07\\ 
& $Q_4$ vs.~$Q_5$ & 10250 & 10251 & 1270 & 1251 & 
0.18 & 16.8 & 12.53 & 34.1 & 0.0001\\ 
\midrule
Economics & $Q_1$ vs.~$Q_2$ & 1462 & 1462 & 283 & 348 & 0.25 & 7.03 & 6.43 & 9.3 & 0.5\\ 
& $Q_2$ vs.~$Q_3$ & 1461 & 1462 & 184 & 203 & 
0.23 & 7.14 & 6.13 & 16.5 & 0.3\\ 
& $Q_3$ vs.~$Q_4$ & 1461 & 1466 & 151 & 155 & 
0.2 & 9.67 & 6.51 & 48.6 & 0.005\\ 
& $Q_4$ vs.~$Q_5$ & 1462 & 1461 & 135 & 135 & 
0.15 & 19.66 & 14.89 & 32.0 & 0.3\\ 
\midrule
Mathematics & $Q_1$ vs.~$Q_2$ & 2339 & 2340 & 328 & 451 & 0.29 & 7.61 & 7.31 & 4.1 & 0.8\\ 
& $Q_2$ vs.~$Q_3$ & 2340 & 2339 & 192 & 209 & 
0.21 & 9.97 & 7.07 & 40.9 & 0.04\\ 
& $Q_3$ vs.~$Q_4$ & 2340 & 2340 & 130 & 144 & 
0.22 & 14.74 & 9.85 & 49.6 & 0.03\\ 
& $Q_4$ vs.~$Q_5$ & 2339 & 2340 & 119 & 130 & 
0.23 & 17.99 & 11.66 & 54.4 & 0.009\\ 
\midrule
Biology & $Q_1$ vs.~$Q_2$ & 37516 & 37504 & 10603 & 13261 & 0.32 & 18.93 & 15.33 & 23.5 & 7.1e-19\\ 
& $Q_2$ vs.~$Q_3$ & 37492 & 37518 & 10628 & 10847 & 0.26 & 19.59 & 16.71 & 17.3 & 1.7e-10\\ 
& $Q_3$ vs.~$Q_4$ & 37501 & 37507 & 10216 & 10346 & 0.25 & 22.64 & 19.0 & 19.1 & 1.0e-12\\ 
& $Q_4$ vs.~$Q_5$ & 37503 & 37506 & 11228 & 10427 & 0.27 & 31.43 & 23.55 & 33.5 & 2.3e-35\\ 
\midrule
Comp. Sci. & $Q_1$ vs.~$Q_2$ & 10701 & 10702 & 1890 & 2722 & 0.3 & 7.14 & 5.39 & 32.6 & 2.4e-05\\ 
& $Q_2$ vs.~$Q_3$ & 10702 & 10701 & 1381 & 1515 & 
0.23 & 8.87 & 6.66 & 33.1 & 0.0002\\ 
& $Q_3$ vs.~$Q_4$ & 10701 & 10702 & 1109 & 1177 & 
0.22 & 12.22 & 11.53 & 6.0 & 0.6\\ 
& $Q_4$ vs.~$Q_5$ & 10702 & 10701 & 1041 & 1004 & 
0.21 & 17.74 & 13.79 & 28.7 & 0.006\\ 
\midrule
Geology & $Q_1$ vs.~$Q_2$ & 1257 & 1258 & 86 & 110 & 0.27 & 13.74 & 10.99 & 25.0 & 0.3\\ 
& $Q_2$ vs.~$Q_3$ & 1257 & 1259 & 47 & 46 & 
0.15 & 13.9 & 17.01 & -18.3 & 0.5\\ 
& $Q_3$ vs.~$Q_4$ & 1257 & 1258 & 44 & 38 & 
0.12 & 19.04 & 14.43 & 31.9 & 0.2\\ 
& $Q_4$ vs.~$Q_5$ & 1257 & 1258 & 44 & 48 & 
0.12 & 24.77 & 25.2 & -1.7 & 0.9\\ 
\midrule
Psychology & $Q_1$ vs.~$Q_2$ & 6494 & 6492 & 943 & 1599 & 0.31 & 15.27 & 11.08 & 37.8 & 9.3e-05\\ 
& $Q_2$ vs.~$Q_3$ & 6489 & 6494 & 548 & 611 & 
0.23 & 16.61 & 12.83 & 29.5 & 0.004\\ 
& $Q_3$ vs.~$Q_4$ & 6493 & 6493 & 451 & 466 & 
0.18 & 19.99 & 18.26 & 9.5 & 0.3\\ 
& $Q_4$ vs.~$Q_5$ & 6489 & 6495 & 365 & 377 & 
0.17 & 32.1 & 24.56 & 30.7 & 0.01\\ 
\midrule
Engineering & $Q_1$ vs.~$Q_2$ & 814 & 820 & 65 & 98 & 0.29 & 4.4 & 1.8 & 144.3 & 0.08 \\ 
& $Q_2$ vs.~$Q_3$ & 818 & 818 & 44 & 48 & 
0.19 & 3.13 & 3.19 & -1.9 & 0.9\\ 
& $Q_3$ vs.~$Q_4$ & 817 & 818 & 22 & 25 & 
0.26 & 5.86 & 4.71 & 24.2 & 0.6\\ 
& $Q_4$ vs.~$Q_5$ & 816 & 819 & 29 & 25 & 
0.15 & 5.7 & 11.48 & -50.3 & 0.2\\
\midrule
\end{tabular}
\end{center}
}}
\end{table}

\end{minipage}

\clearpage
{\noindent\Large \textbf{Supplementary References}}

\bibliography{naturebib}
\bibliographystyle{naturemag}